\documentclass[11pt,a4paper]{article}
\usepackage{graphicx}
\usepackage[bf,hang,margin=7truemm]{caption}[2007/04/09]
\usepackage{amsmath}
\usepackage{amssymb}
\usepackage{bm}
\usepackage{url}
\def\fige#1#2#3{
\begin{figure}[!tp]
\begin{center}
\includegraphics[width=#3\textwidth]{#1.eps}
\caption{#2}
\label{fig:#1}
\end{center}
\end{figure}}
\def\figeb#1#2#3{
\begin{figure}[!b]
\begin{center}
\includegraphics[width=#3\textwidth]{#1.eps}
\caption{#2}
\label{fig:#1}
\end{center}
\end{figure}}
\begin{document}
\title{\rightline{\normalsize KUNS-2400}
\vspace{20pt}
\sf\LARGE Equilibrium Distribution of Labor Productivity:\\ A Theoretical Model}
\author{
{\large Hideaki Aoyama}\footnote{hideaki.aoyama@scphys.kyoto-u.ac.jp}\\
\normalsize Department of Physics, Graduate School of Science, Kyoto University,\\
Kyoto 606-8502, Japan
\\[8pt]
{\large Hiroshi Iyetomi and Hiroshi Yoshikawa}\\
\normalsize Faculty of Economics, University of Tokyo,\\
Tokyo 113-0033, Japan}
\maketitle
\begin{abstract}
We construct a theoretical model for equilibrium distribution of workers across sectors with different labor productivity, assuming that a sector can accommodate a limited number of workers which depends only on its productivity. A general formula for such distribution of productivity is obtained, using the detail-balance condition necessary for equilibrium in the Ehrenfest-Brillouin model. We also carry out an empirical analysis on the average number of workers in given productivity sectors on the basis of an exhaustive dataset in Japan.
The theoretical formula succeeds in explaining the two distinctive observational facts in a unified way, that is, a Boltzmann distribution with negative temperature on low-to-medium productivity side and a decreasing part in a power-law form on high productivity side.
\end{abstract}
\newpage
\section{Introduction}

The concept of equilibrium plays a central role in economics. Of all, the most influential is the Walrasian general equilibrium as represented by Arrow and Debreu~\cite{arrow1954}. Though it is a grand concept, and well established in the profession, it cannot be more different from the real economy. The Walrasian theory specifies preferences and technologies of all the consumers and firms, and defines the equilibrium in which micro-behaviors of all the economic agents are precisely determined. It is just as one analyses object such as gas comprising many particles by determining the equations of motion for all the particles. Physicists know that this approach though it may look reasonable at first sight, is actually infeasible and on the wrong track. Instead, following the lead of Maxwell, Boltzmann and Gibbs, they have developed statistical physics. Curiously, despite of the fact that the macroeconomy consists of many heterogeneous consumers and firms, the basic method of statistical physics has had almost no impact on economics.

In the Walrasian equilibrium, the marginal productivities of production factors such as labor and capital are equal in all the firms, industries, and sectors. Otherwise, there exists inefficiency in the economy, and it contradicts the notion of equilibrium. However, in the real economy, we actually observe significant productivity dispersion. That is, there is a distribution rather than a unique level of productivity.
Search theory has attempted to explain such distribution by considering frictions and search costs which exist in the real economy~\cite{diamond2011,mortensen2011,pissarides2011}. However, it is still based on representative agent assumptions~\cite{yoshikawaUT2011}. 

To explain equilibrium distribution, the most natural and promising approach is to eschew pursuit of precise micro-behavior on representative agent assumptions, and resort to the method of statistical physics. Foley~\cite{foley1994} is a seminal work which applies such statistical method to the general equilibrium model. Yoshikawa~\cite{yoshikawa2003} argues that the study of productivity dispersion provides correct micro-foundations for Keynesian economics, and that to explain distribution of productivity we should apply the method of statistical physics. In a series of papers, we have attempted to establish the empirical distribution using a large data set covering more than a million firms in the Japanese manufacturing and non-manufacturing industries~\cite{fai2009superstatistics,souma2009distribution,aoyama2010real,aoyama2010productivity}. To explain this empirically observed distribution of productivity, Iyetomi~\cite{iyetomiptp2012} introduced the notion of \textit{negative temperature}. Based on this notion of negative temperature, Yoshikawa~\cite{yoshikawaUT2011} also made a similar attempt with the help of grandcanonical partition function.

In this paper, we explore the problem from a different angle than the standard entropy maximization. Before doing so, we first update our empirical investigation of distribution of labor productivity in section 2. Most of theoretical works exploring distribution of productivity resort to the straight-forward entropy maximization. Instead, Scalas and Garibaldi~\cite{gsCUP} suggest that we study the same problem using the Ehrenfest-Brillouin model, a Markov chain which describes random creations and destructions in system comprising many elements moving across a finite number of categories. Following their lead, we present such a model in section 3. By considering detailed balance, we derive the stationary distribution of the model which explains the empirically observed distribution.  Section 4 offers brief concluding remarks.

\section{Distribution of Labor Productivity}
The labor productivity denoted by $c$, is simply defined by
\begin{equation}
c:=\frac{Y}{n}.
\end{equation}
Here, $Y$ is the value added in units of $10^3$ yen, and $n$ the number of workers.
Iyetomi~\cite{iyetomiptp2012} has studied the firm data in Japan for the year 2006.
Since then, we have obtained the data up to the year 2010 and will use
the 2008 data in this paper, as it contains the largest number of firms.

Let us briefly review the method of calculating the value added $Y$.
Dataset is constructed by unifying two datasets, the Nikkei Economic Electric Database (NEEDS) \cite{nn} for large firms (most of which are listed) and
and the Credit Risk Database (CRD) \cite{crd} for small to medium firms.
The value added $Y$ is calculated by the so-called BoJ method, established by the Statistics Department of the Bank of Japan, and gives the value added as the sum of net profits, labor costs, financing costs, rental expenses, taxes, and depreciation costs.
Although the original datasets contain over a million firms together, by limiting the analysis to firms which have non-empty entries in all these items, we end up with 180,181 firms for 2008.%
\footnote{This covers all the industrial sectors, except for 
the finance and insurance (95 firms in 2008), 
the deep-sea foreign transport of freight (9 firms in 2008), 
and the holding companies (138 firms in 2008), all of which show abnormally high
value of the productivity compared with other sectors.}

\newcommand{\mn}{\bar{n}}

\fige{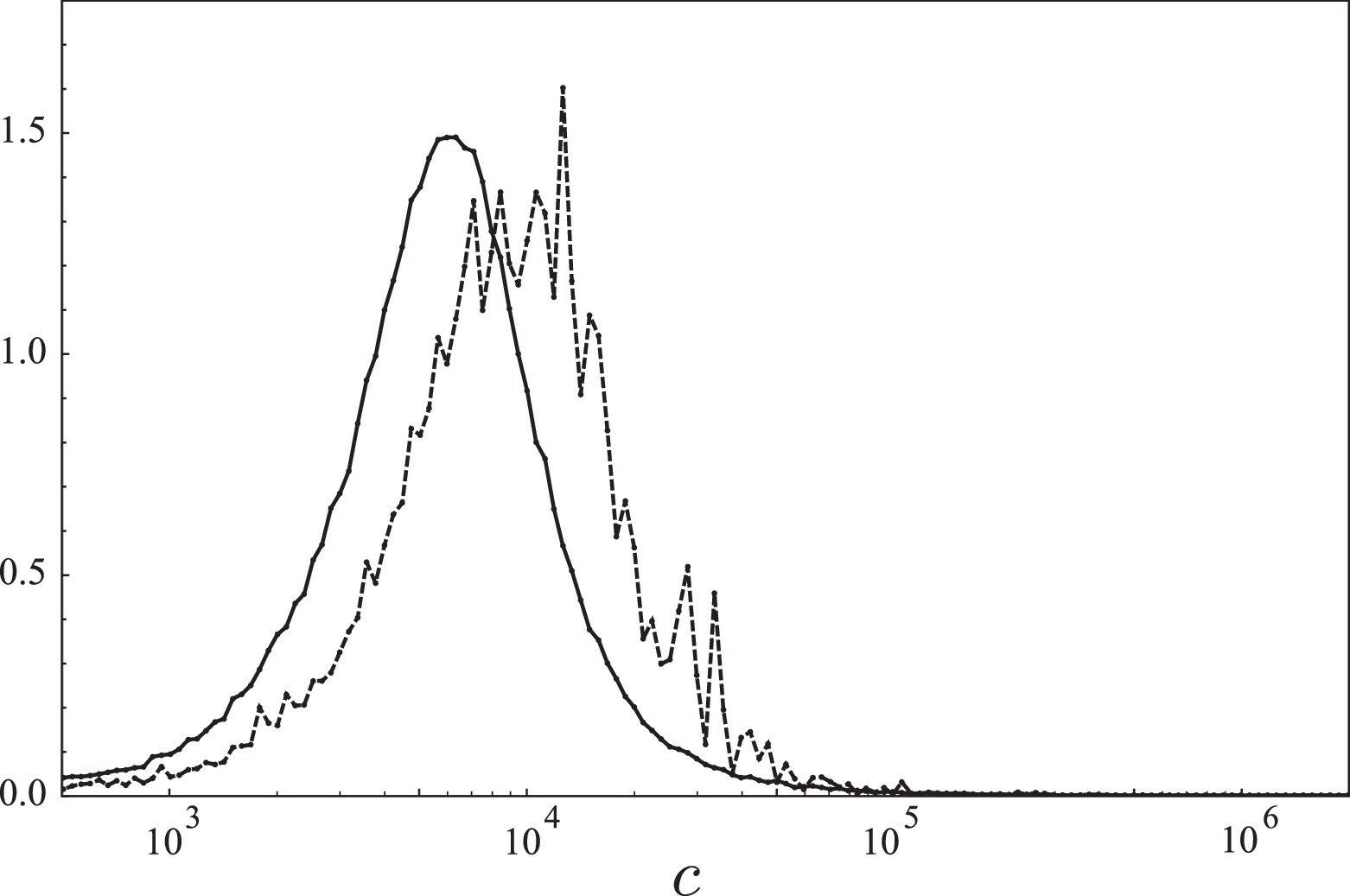}{PDF of the firms' $\log c$ (solid lines) and workers' $\log c$ (dashed lines).
The productivity $c$ is in units of $10^3$ yen/person.}{0.7}

Figure \ref{fig:3_pdf0_2008} shows the PDF of the firms' and the worker's $\log$ of
the labor productivity $c$ in units of $10^3$ yen/person. The fact that the major peak of the latter is shifted to right compared to that of the former indicates that the average number of workers per firm $\mn$ increases in this region.
In fact, Fig.~\ref{fig:3_nbar0_2008} shows the dependence of 
$\mn$ on the labor productivity of the firm ($c$).
We observe that as the productivity rises, it first goes up to about $n\simeq200$
and then decreases. Iyetomi~\cite{iyetomiptp2012} explained the upward-sloping distribution in the low productivity region by introducing the negative temperature theory.
The downward-sloping part in the high productivity region is close to linear (denoted by the dotted line) 
in this double-log plot. This indicates that it obeys the power low:
\begin{equation}
\mn \propto c^{-\gamma}.
\label{ncgamma}
\end{equation}

\fige{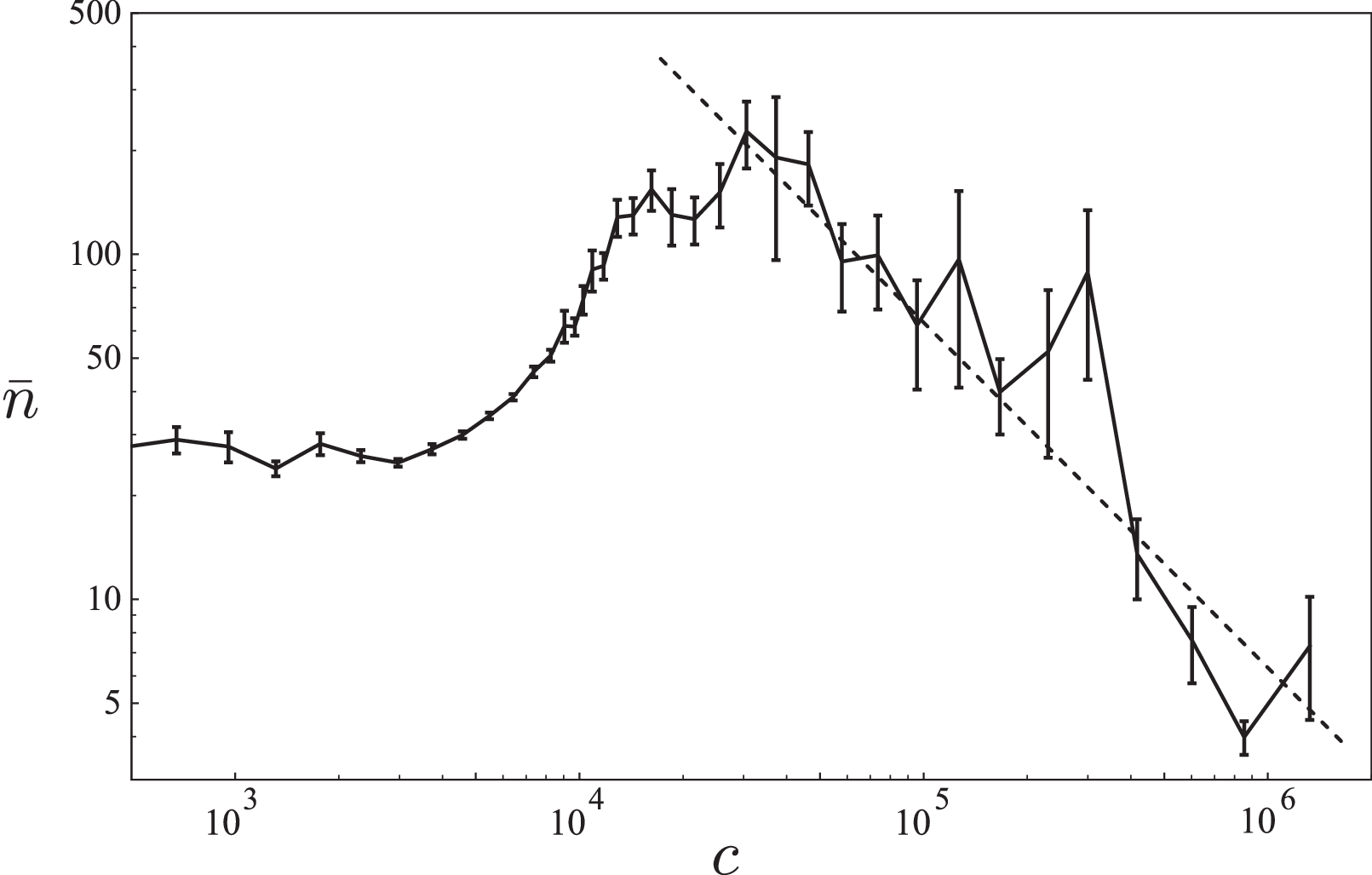}{Dependence of the average number of workers $\mn$
on the labor productivity $c$ of firms (dots connected by thick lines)
The dotted straight line has the gradient $-1$, that is, $\mn\propto 1/c$.}{0.8}

We have studied this phenomenon in the period of 2000 through 2008, and not only for all the sectors but also for the manufacturing and the non-manufacturing sectors separately. It turns out that we  
always find the qualitatively same pattern as shown in Fig.~\ref{fig:3_nbar0_2008}; the number of workers exponentially increases as $c$ increases up to a certain level of productivity whereas it decreases following power law (Eq.(\ref{ncgamma})) in the high productivity region. We thus conclude that this broad shape of distribution of productivity among firms is quite robust and universal.
We note that this is somewhat counter-intuitive in the sense that
firms that achieved higher productivity through innovation and high-quality
management would grow larger, so that equilibrium distribution would simply
have monotonically increasing $\mn$ with $c$.
Therefore we need to find what is the main reason that causes this behavior, which 
we will do in the following section.

\section{The Ehrenfest-Brillouin Model}
\newcommand{\pa}{P(i,j;k,\ell)}
\newcommand{\pb}{P(k,\ell;i,j)}
\newcommand{\na}{N(i,j;k,\ell)}
\newcommand{\nb}{N(k,\ell;i,j)}
One way to analyze the equilibrium distribution of labor productivity based on statistical physics is to maximize entropy. Instead, Scalas and Garibaldi~\cite{gsCUP} suggest that we can usefully apply the Ehrenfest-Brillouin model, a Markov chain to analyze the problem. In this section, we present such a model. 

The macroeconomy consists of many firms with different levels of productivity. Differences in productivity arise from different capital stocks, levels of technology and/or demand conditions facing firms. We call a group of firms with the same level of productivity a \textit{cluster}.
Workers randomly move from a cluster to another for various reasons at various times. Despite of these random changes, the distribution of labor productivity as a whole remains stable because those incessant random
movements balance with each other.
This balancing must be achieved for each cluster, and is called detail-balance.
In the following, we present a general treatment of this detail-balance using particle-correlation theory \`a la \cite{costantini1989classical}. In doing so, we make an assumption that the number of workers who belong to clusters with high productivity is constrained; see also Ref.~\cite{yoshikawaUT2011}.

We denote the number of workers who belong to a cluster with the level of labor productivity, $c_{i}$ by $n_{i}$. The total output in the economy as a whole, $Y$ is assumed to be equal to aggregate demand $D$, and is given:
\begin{equation}
Y=\sum_{i}c_{i}n_{i} = D
\label{con0}
\end{equation}

For the productivity distribution of workers specified by $\{n_1, n_2 ,n_3, \dots\}$
to be in equilibrium, the number of workers who move {\it out} of cluster $i$
per unit time must be equal to that of workers who move
{\it into} this cluster per unit time.
We consider the minimal process that satisfies the condition that the total output in the economy as a whole is conserved with Eq. (\ref{con0}).
In this process, two workers move simultaneously \cite{gsCUP,scalasg}.
This is illustrated in Figure \ref{fig:process}(a):
A worker in a cluster with productivity $c_i$
and a worker in a cluster with productivity $c_j$
move to clusters with productivities $c_k$ and $c_\ell$, respectively. For the total output $Y$ to remain constant, the following condition must be satisfied:
\begin{equation}
c_i+c_j=c_k+c_\ell.
\label{con1}
\end{equation}

\figeb{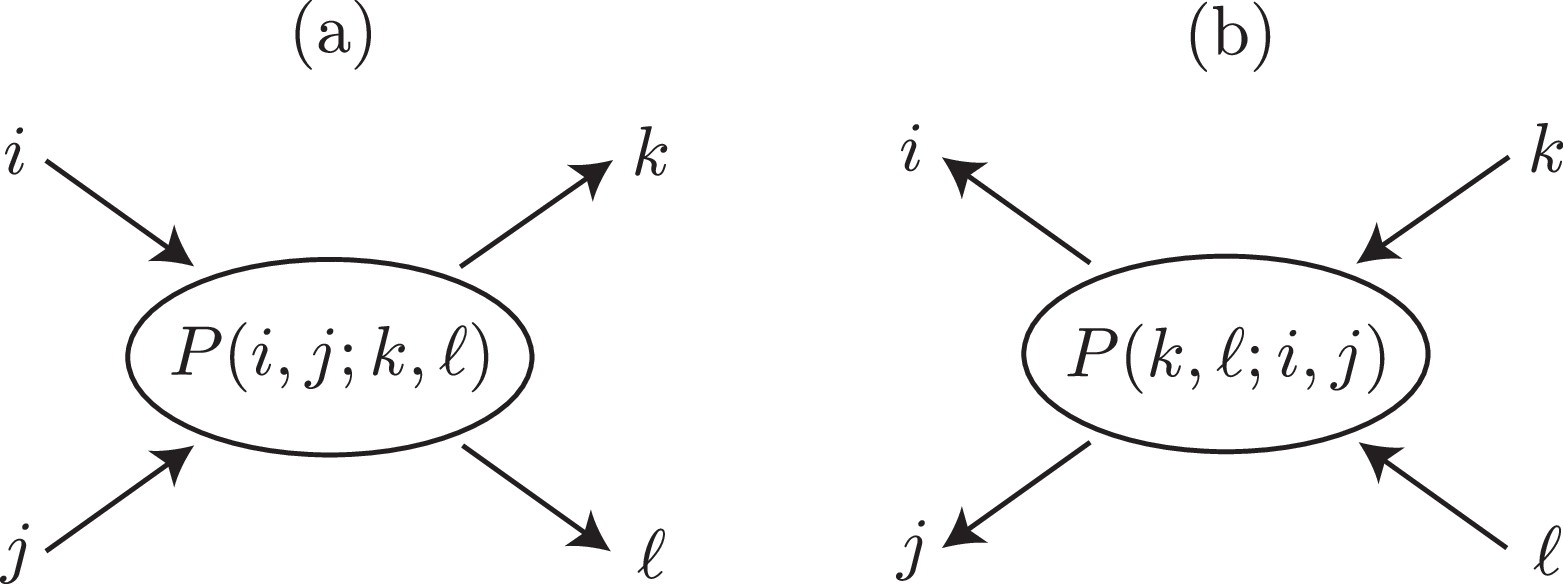}{Elementary process where (a) a worker at the firm $i$ and a worker
at the firm $j$ move to firms $k$ and $\ell$, with probability $\pa$
and (b) the reverse process with probability $\pb$.}{0.6}

Such job-switchings occur for various unspecifiable reasons. The best we can do is to consider a Markov chain defined by transition rates, $\pa$.
They have the following trivial symmetries:
\begin{equation}
P(i,j;k,\ell)=P(j,i;k,\ell), \quad
P(i,j;k,\ell)=P(i,j;\ell,k).
\label{tsym}
\end{equation}
We also assume that the reverse process, illustrated in Figure \ref{fig:process}(b) 
occurs with the same probability:
\begin{equation}
\pa=\pb.
\label{timeref}
\end{equation}

Equilibrium condition then requires
the number of workers moving from $(i,j)$ to $(k,\ell)$ per unit time denoted by $\na$ must be equal to the number of those from $(k,\ell)$ to $(i,j)$ denoted by $\nb$:
\begin{equation}
\na=\nb.
\label{db}
\end{equation}
The flux $\na$ is proportional to the numbers of workers in clusters $i$ and $j$,
$n_i$ and $n_j$ and the corresponding transition rate $\pa$:
\begin{equation}
\na\propto \pa n_i n_j.
\end{equation}

The fundamental assumption we make is that a cluster with productivity $c$ can accommodate $g(c)$ workers at most.
It means that
\begin{equation}
\na=\pa n_i n_jL(c_k,n_k)L(c_\ell,n_\ell),
\label{np}
\end{equation}
where $L(c,n)$ is a function that limits the number of workers in a cluster with productivity $c$:
\begin{equation}
L(c,n)=0 \quad \mbox{for} \quad n\ge g(c).
\label{lc}
\end{equation}

\newcommand{\tn}{H}
One can obtain a general solution in terms of $L(c,n)$ for the detail-balance equation (\ref{db}) in the following way.
Thanks to Eq.~(\ref{timeref}), substituting Eq.~(\ref{np}) into Eq.~(\ref{db}) enables us to find
\begin{equation}
\tn(c_i)\tn(c_j)= \tn(c_k)\tn(c_\ell),
\label{tn}
\end{equation}
where 
$\tn(c):=n/L(c, n)$.
Because of Eq.~(\ref{con1}), we obtain
\begin{equation}
\tn(c_i)\tn(c_j)=\tn(c_i+c_j)\tn(0)
\end{equation}
or, by denoting $G(c):=\log(\tn(c)/\tn(0))$,
\begin{equation}
G(c_i)+G(c_j)=G(c_i+c_j).
\end{equation}
This proves that $G(c)$ is linear in $c$. It leads us to
\begin{equation}
n=L(c,n)e^{-\beta (c-\mu)},
\label{gsol}
\end{equation}
where $\mu$ and $\beta$ are real free parameters.
Once the function $L(c,n)$ is given, the equilibrium distribution $\mn$ can be obtained by solving the above.

%

\fige{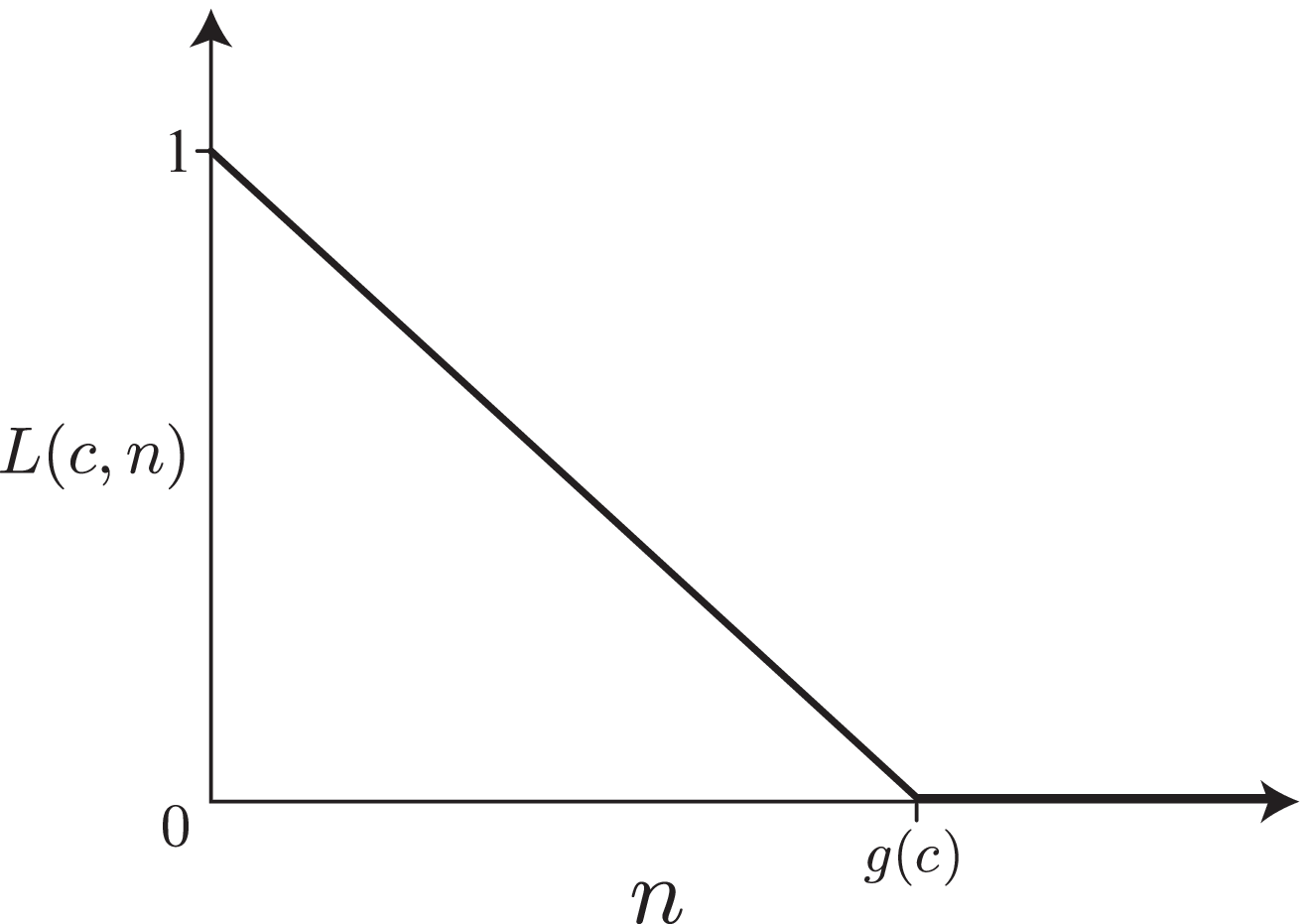}{A model of worker limitation $L(n,c)$ in Eq.~(\ref{eq:Lhype}).}{0.5}
In order to model the distribution of labor productivity, 
we need to allow $g$ to be any integer number.
Furthermore, we find it most natural to choose $L(c,n)$ so that
it is continuous at $n=g(c)$. Here we adopt a simple linear model,
\begin{equation}
L(c,n) = \theta(g(c)-n)\frac{g(c)-n}{g(c)},
\label{eq:Lhype}
\end{equation}
as depicted in Fig.~\ref{fig:Lhype}.
We can reasonably assume that $L(c,0)=1$, because there would be no restrictions for
hiring workers if there are none in the firm.

By substituting Eq.~(\ref{eq:Lhype}) into Eq.~(\ref{gsol}) and solving for $n$, we finally obtain
\begin{equation}
\mn=\frac{g(c)}{g(c)e^{\beta(c-\mu)}+1}.
\label{ggsol}
\end{equation}
This is a simple extension of the Fermi-Dirac statistics. In passing, we note that
the partition function $Z$ that yields Eq.~(\ref{ggsol}) is
\begin{equation}
Z=\left(1+\frac1{g(c)}e^{-\beta(c-\mu)}\right)^{g(c)}.
\end{equation}
It is a reasonable extension of the Fermi-Dirac statistics
in the sense that the partition function has the expansion
\begin{equation}
Z=1+e^{-\beta(c-\mu)}+\cdots
\end{equation}
and yet allows existence of $g(c)$ levels.

\section{Results and Discussion}
First, we note that 
when there is no limit to the number of the workers, {\it i.e.}, $g \rightarrow \infty$, 
Eq.~(\ref{ggsol}) boils down to the Boltzmann distribution,
\begin{equation}
\mn=e^{-\beta(c-\mu)}.
\label{eq:BD}
\end{equation}
When we apply Eq.~(\ref{eq:BD}) to low-to-intermediate range of $c$ where
$\mn$ is an exponentially increasing function of $c$ as observed in Fig.~\ref{fig:3_nbar0_2008}, we must have
\begin{equation}
\beta<0,
\end{equation}
the negative temperature. We, therefore, assume that $\beta$ is negative in the following. The current model thus incorporates the Boltzmann statistics model with negative temperature advanced in Ref.~\cite{iyetomiptp2012}. 

Secondly, we recall the observation that the power law (\ref{ncgamma}) holds for $\mn$ in the high productivity side. We can use this empirical fact to determine the functional form for $g(c)$. Equation (\ref{ggsol}) implies that when temperature is negative, $\mn$ approaches $g(c)$ in the limit $c \rightarrow \infty$.
These arguments persuade us to adopt the following anzatz for $g(c)$,
\begin{equation}
g(c)=A c^{-\gamma}.
\label{eq:gc}
\end{equation}

Given the present model, explaining the empirically observed distribution of productivity is equivalent to determining four parameters, $\beta$, $\mu$, $A$, and $\gamma$ in Eqs.~(\ref{ggsol}) and (\ref{eq:gc}). We estimate these four parameters by the $\chi^2$ fit to
the empirical results as shown in Fig.~\ref{fig:3_nbar0_2008}.

Figure~\ref{fig:5_pw_fitplot_2008} demonstrates the results of the best fit for three datasets of firms, namely, those in all the sectors, the manufacturing sector, and the non-manufacturing sector. The fitted parameters are listed in Table~\ref{tab:bf}. The present model is quite successful in unifying the two opposing functional behaviors of the average number of workers in the low-to-medium and high productivity regimes. The crossover takes place at the productivity $c_{\rm p}$ in the nonmanufacturing sector which is about 40\% as high as that in the manufacturing sector.

Also we see that the inverse temperature $\beta$ of the non-manufacturing sector is just half of that of the manufacturing sector. This manifests there is a much wider demand gap in the non-manufacturing sector. The economic system is thus far away from equilibrium in demand exchange. In contrast, $\beta$ times $\mu$ gives almost the same values for the two sectors, indicating that the system seems to be in equilibrium as regards exchange of workers. These findings agree well with those obtained in the previous study~\cite{iyetomiptp2012}.

\fige{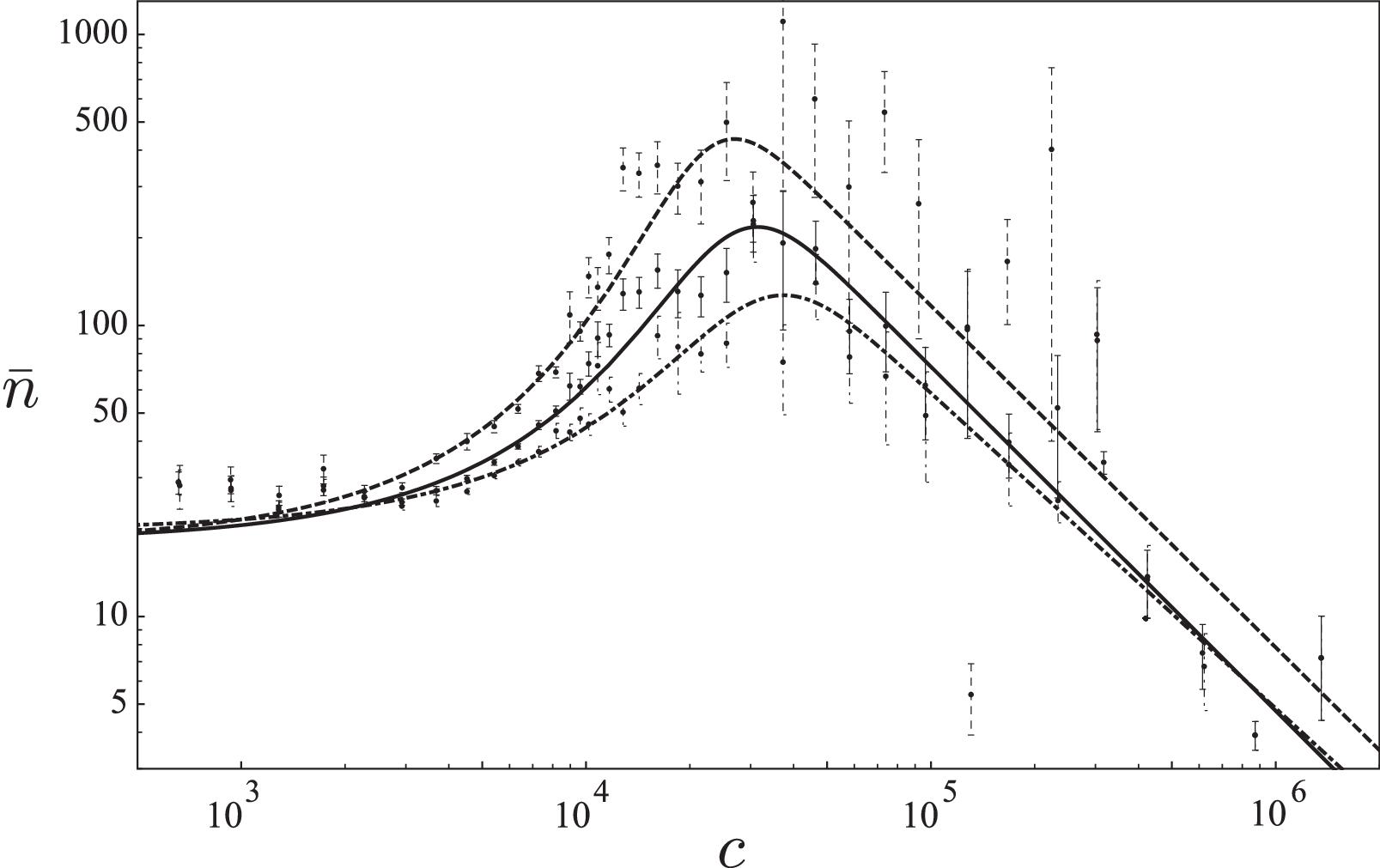}{The best fits to the data; Solid curve is for all
sectors, the dashed curve for the manufacturing sector, the dot-dashed curve for
the non-manufacturing sector.}{0.8}

\begin{table}
\begin{center}
\begin{tabular}{c|rrrrr}
&$\beta\ (\times 10^{-4})$&$\mu\ (\times 10^4)$&$A\ (\times 10^7)$
&\hspace{15pt}$\gamma$\hspace{15pt}
&$c_{\rm p}\ (\times 10^4)$\\
\hline\hline
All&$-1.25$&$-2.32$&5.84&1.18&3.14\\
Manucacturing&$-1.78$&$-1.63$&8.51&1.17&2.70\\
Non-manufacturing&$-0.86$&$-3.47$&1.52&1.08&3.74
\end{tabular}
\caption{Best-fit parameters and the position of the peak $c_{\rm p}$.}
\label{tab:bf}
\end{center}
\end{table}

\section{Conclusion}
A theoretical model was proposed to account for empirical facts on distribution of workers across clusters with different labor productivity. Its key idea is to assume that there are restrictions on capacity of workers for clusters with high productivity. This is a rational assumption because most of firms belonging to such superb clusters are expected to be in a cutting-edge stage to lead industry. We then derived a general formula for the equilibrium distribution of productivity, adopting the Ehrenfest-Brillouin model along with the detail-balance condition necessary for equilibrium. Fitting of the model to the empirical results confirmed that the theoretical model could encompass both of a Boltzmann distribution with negative temperature on low-to-medium productivity side and a decreasing part in a power-law form on high productivity side.

\subsection*{Acknowledgments}
The authors would like to thank 
Yoshi Fujiwara, Yuichi Ikeda and Wataru Souma
for helpful discussions and comments during the course of this work.
and Toshiyuki Masuda, Chairman of The Kyoto Shinkin Bank for his comments
on Japanese small to medium firms.
We would also thank the Credit Risk Database for the data used in this paper.
This work is supported in part by {\it the Program for Promoting Methodological 
Innovation in Humanities and Social Sciences by Cross-Disciplinary Fusing} of 
the Japan Society for the Promotion of Science.

\end{document}